\documentclass[aps,pra,reprint,showpacs,longbibliography,nofootinbib,superscriptaddress]{revtex4-1}

\usepackage{tikz}
\usepackage{pgfplots}
\usetikzlibrary{plotmarks}
\usetikzlibrary{arrows}
\tikzset{every text node part/.style={font=\footnotesize},>=latex'} %global tikz settings go here.
\usetikzlibrary{calc} % tikz coordinate calculations

\usepackage{hyperref}
\hypersetup{
	bookmarksopen=true,%
	bookmarksnumbered=true,%
	colorlinks=true,%
    	linkcolor=blue,          % color of internal links
    	citecolor=blue,        % color of links to bibliography
    	filecolor=blue,      % color of file links
    	urlcolor=blue,           % color of external links
	pdfstartview=FitH,%
	pdfnewwindow=true% links in new window
}

\usepackage{amsmath} 
\usepackage{amssymb}

\usepackage{mathtools} \mathtoolsset{showonlyrefs}

\usepackage{txfonts}

\usepackage{microtype}

\newcommand{\bra}[1]{\langle#1\rvert}
\newcommand{\ket}[1]{\lvert#1\rangle}
\newcommand{\qprod}[2]{ \langle #1 | #2 \rangle }
\newcommand{\braopket}[3]{\langle #1 | #2 | #3\rangle}
\newcommand{\intk}{\int \dif k \,}
\newcommand{\Vt} {\tilde{V}}
\newcommand{\spi}{\frac{1}{\sqrt{2 \pi} }}
\newcommand{\expect}[1]{ \langle #1 \rangle}

\DeclareMathOperator{\ain}{\mathit{a}_\text{in}}
\DeclareMathOperator{\aind}{\mathit{a}_\text{in}^\dag}
\DeclareMathOperator{\aout}{\mathit{a}_\text{out}}
\DeclareMathOperator{\aoutd}{\mathit{a}_\text{out}^\dag}

\DeclareMathOperator{\aRin}{\mathit{r}_\text{in}}
\DeclareMathOperator{\aRind}{\mathit{r}_\text{in}^\dag}
\DeclareMathOperator{\aLin}{\mathit{\ell}_\text{in}}

\DeclareMathOperator{\aRout}{\mathit{r}_\text{out}}

\DeclareMathOperator{\aLout}{\mathit{\ell}_\text{out}}

\newcommand{\aR}[1]{\mathit{r}_{#1}}
\newcommand{\aRd}[1]{\mathit{r}_{#1}^\dag}
\newcommand{\aL}[1]{\mathit{\ell}_{#1}}
\newcommand{\aLd}[1]{\mathit{\ell}_{#1}^\dag}
\DeclareMathOperator{\aRinout}{\mathit{r}_\text{in/out}}
\newcommand{\wR}{\omega_r}
\newcommand{\wL}{\omega_\ell}

\DeclareMathOperator{\azind}{\mathit{\mathring{a}}_\text{in}^\dag}
\DeclareMathOperator{\azout}{\mathit{\mathring{a}}_\text{out}}

\DeclareMathOperator{\ainout}{\mathit{a}_\text{in/out}}
\DeclareMathOperator{\azinout}{\mathit{\mathring{a}}_\text{in/out}}

\newcommand{\me}{\mathrm{e}}
\newcommand{\mi}{\mathrm{i}}

\newcommand{\dif}{\mathrm{d}}
\begin{document}
\title{Input-Output Formalism For Few-Photon Transport in \\One-Dimensional Nanophotonic Waveguides Coupled to a Qubit}
\author{Shanhui Fan}
\email{shanhui@stanford.edu}
\author{ \c{S}\"ukr\"u Ekin Kocaba\c{s}}
\affiliation{Ginzton Laboratory, Department of Electrical Engineering, Stanford University, Stanford, CA 94305}
\author{Jung-Tsung Shen}
\affiliation{Department of Electrical \& Systems Engineering, Washington University, St. Louis, MO 63130}
\date{November 2010}

\begin{abstract}
We extend the input-output formalism of quantum optics to analyze few-photon transport in waveguides with an embedded qubit. We provide explicit analytical derivations for one and two-photon scattering matrix elements based on operator equations in the Heisenberg picture.
\end{abstract}

% 03.65.Nk 	Scattering theory 
% 32.50.+d 	Fluorescence, phosphorescence (including quenching)
% 42.50.Ct 	Quantum description of interaction of light and matter; related experiments 
% 42.50.Pq 	Cavity quantum electrodynamics; micromasers 

\maketitle

%%%%%%%%%%%%%%%%%%%%%%%%%
%%%%%%%%%%%%%%%%%%%%%%%%%
\section{Introduction}

In the context of quantum information technology, including quantum computing devices, understanding the interaction between a few-photon state and a two-level atom plays an important role \cite{OBrien2009,Kimble2008,Schoelkopf2008}. The photons are a possible candidate for the `flying qubit' that carries the information, and the two-level atom constitutes the `stationary qubit' where the flying qubits are generated on demand and correlated with each other.

Recently, there has been an increased activity in analyzing the properties of photons propagating in a waveguide coupled to a qubit---a two-level quantum mechanical system. Experimental demonstration of the control of single photons was made in a waveguide coupled to an optical cavity with an atom in its near field \cite{Dayan2008}. Similar effects were observed in the microwave domain, when low frequency photons in a transmission line were coupled to a superconducting qubit \cite{Wallraff2004,Astafiev2010}, which later was shown to act as a photon amplifier \cite{Astafiev2010a}.

To theoretically model such systems one needs to consider a continuous set of waveguide modes that are free to propagate in one dimension, either directly coupled to a multi-level system (referred to as an `atom' in the paper), or indirectly coupled through an optical cavity with a discrete set of modes. Photon transport properties are non-trivial in these structures \cite{Kojima2003,Shen2005,Shen2007,Shen2007a} which can be tailored to perform logic operations \cite{Chang2007} or form a diode \cite{Roy2010}. Exact solutions of one and two-photon scattering have first been reported in \cite{Shen2005,Shen2007a}.

The most widely used theoretical approach is to treat the set of equations in the Schr\"odinger picture, and apply the Lippmann-Schwinger formalism to calculate the reflection and transmission properties of the single and multi-photon states \cite{Shen2007,Yudson2008,Witthaut2010,Zheng2010,Liao2010}. An alternative technique is to use the reduction formulas from field theory to calculate the scattering matrix of the system \cite{Shi2009,Shi2010}. Time-domain simulations that take the waveguide dispersion into account are also possible, and an interesting radiation trapping mechanism was recently predicted \cite{Longo2010}.

In this paper, we extend the input-output formalism \cite{Gardiner1985,[{}][{. See Chap. 7 for the input-output formalism in cavities, and Sec. 10.5 for resonance fluorescence.}]Walls2008} of quantum optics---an Heisenberg picture approach originally introduced to analyze the interaction between an atom in a cavity and a continuous set of electromagnetic states outside of the atom-cavity system---to analyze the transport of few-photon states in a waveguide with an embedded qubit. In the input-output formalism one obtains a \textit{nonlinear} set of operator equations based on the Hamiltonian of the system. For a coherent or a squeezed state input, this formalism has been extensively used to calculate various coherence properties of the output state of light. Here, we show that one can adopt this formalism to obtain exact results regarding one or two photon properties. To do so, we establish a relationship between the input-output formalism and the scattering matrix elements of the system. Our approach complements the existing theoretical literature and bridges different analytical techniques. 

The paper is organized as follows: In Section \ref{sec.hamiltonian} we introduce the Hamiltonian of the system. In Section \ref{sec.connection} we build the link between the scattering theory and the input-output formalism and continue in Section \ref{sec.onePhoton} with the derivation of the one-photon transport properties. In Section \ref{sec.twoPhoton} we show how to extend the calculations to the two-photon case. In Section \ref{sec.coherent} we make observations on correlation function calculations based on coherent state inputs and end with our conclusions in Section \ref{sec.conclusion}.

%%%%%%%%%%%%%%%%%%%%%%%%%
%%%%%%%%%%%%%%%%%%%%%%%%%
\section{Hamiltonian} \label{sec.hamiltonian}
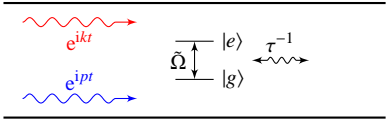
\begin{figure}[tb]
\begin{center}
\begin{tikzpicture}[scale=2.54]>=latex',every text node part/.style={font=\footnotesize}]

%\draw (0,0) node[minimum height=2in,minimum width=3.4in,draw] {$3.4\text{in} \times 2\text{in}$};

\draw[thin] (-0.1,0.1) -- (0.1,0.1) node[right] {$\ket{e}$};
\draw[thin] (-0.1,-0.1) -- (0.1,-0.1) node[right] {$\ket{g}$};
\draw[<->] (0,-0.1) -- (0,0) node[left] {$\tilde{\Omega}$} -- (0,0.1);
\draw [<->,decorate,decoration={snake,amplitude=.4mm,segment length=1.5mm,pre length=2mm,post length=1.5mm}] (0.3,0) -- (0.45,0) node[anchor=south]{$\displaystyle \tau^{-1}$} -- (0.6,0);

\draw [->,red,decorate,decoration={snake,amplitude=.6mm,segment length=3mm,pre length=0mm,post length=1.5mm}] (-0.9,0.2)--(-0.6,0.2) node[anchor=north]{$\displaystyle\me^{\mi k t}$}--(-0.3,0.2);

\draw [->,blue,decorate,decoration={snake,amplitude=.6mm,segment length=3mm,pre length=0mm,post length=1.5mm}] (-0.9,-0.2)--(-0.6,-0.2) node[anchor=south]{$\displaystyle\me^{\mi p t}$}--(-0.3,-0.2);

\draw[thick] (-1,-0.3) -- (1,-0.3) (-1,0.3) -- (1,0.3);
\end{tikzpicture}
\caption{(Color online) Schematic representation of two photons in a waveguide, at frequencies $k$ and $p$, moving to the right, towards a two level atom with energy levels $\ket{g}$ and $\ket{e}$. $\tilde{\Omega}$ is the separation between the energy levels. Coupling of the two level atom to the modes in the waveguide is proportional to $\tau^{-1}$. Long horizontal lines denote the waveguide geometry.}
\label{figGeometry}
\end{center}
\end{figure}

We start by discussing the model Hamiltonian that we will use in this paper. As an illustration of  the formalism, we consider a two-level atom coupled to a single polarization, single-mode waveguide \cite{Shen2005}, and treat the transport properties of few-photon states in such a system (Fig \ref{figGeometry}). The Hamiltonian, $\tilde{H}$, is defined as ($\hbar=1$)
\begin{equation}
\tilde{H}= \tilde{H}_0 + \tilde{H}_1.
\end{equation}
Here $\tilde{H}_0$ describes a chiral, i.e. one-way, waveguide where photons propagate in only one direction
\begin{equation}
\tilde{H}_0 = \int_0^\infty \dif\beta \, \tilde{\omega}(\beta) \, \tilde{a}^\dag_{\beta} \tilde{a}_{\beta}
\end{equation}
and $\tilde{a}_{\beta}$, $\tilde{a}^\dag_{\beta}$ are the annihilation and creation operators for the photons with a wavevector $\beta$ respectively. In Appendix \ref{Appendix.TwoWay} we calculate the reflection and transmission probabilities for photons in a waveguide where the fields propagate in both directions and show that the results are straightforward extensions of the chiral case. The operators obey the commutation relation $[\tilde{a}_\beta, \tilde{a}_{\beta'}^\dag]=\delta(\beta-\beta')$. $\tilde{H}_1$ describes the atom as well as the atom-waveguide interaction
\begin{equation}
\tilde{H}_1 = \frac{1}{2} \tilde{\Omega} \sigma_z + V \int_0^\infty \dif\beta \, \left( \sigma_+ \tilde{a}_\beta + \tilde{a}^\dag_\beta \sigma_- \right).
\end{equation}
Here, $\tilde{\Omega}$ is the atomic transition frequency, $\sigma_\pm$ are the raising and lowering operators for the two level atom and $\sigma_z=2\sigma_+\sigma_- -1$. $V$ denotes the coupling strength between the atomic states and the waveguide modes. The derivation of the Hamiltonian is based on the dipole and the rotating wave approximations \cite{[][{. See Chap. 6. for the derivation of the atom-field Hamiltonian, and Chap. 10 for resonance fluorescence.}]Scully2008} as well as taking the continuum limit for field operators. The details of taking the continuum limit are discussed in Appendix~ \ref{Appendix.Continuum}. 

\begin{figure}[tb]
\begin{center}
\begin{tikzpicture}[]>=latex',every text node part/.style={font=\footnotesize}]

%\draw (0,0) node[minimum height=2in,minimum width=3.4in,draw] {$3.4\text{in} \times 2\text{in}$};

\draw[thin,->] (-0.2in,0) -- (1.7in,0) node[right] {$\beta$};
\draw[thin,->] (0,-0.2in) -- (0,1in) node[above] {$\tilde{\omega}(\beta)$};
\draw[semithick] plot[domain=0:3.14] (\x,{(1-exp(-\x))*2});
\draw[dashed] (0.5in,{(1-exp(-0.5*2.54))*2}) -- (0.5in,0) node[below] {$\displaystyle\beta_0$};
\draw[dashed] (0.5in,{(1-exp(-0.5*2.54))*2}) -- (0in,{(1-exp(-0.5*2.54))*2}) node[left]{$\displaystyle\omega_0$};
\draw[color=red,semithick] plot[domain={2.54*0.1:0.9*2.54}](\x,{(\x - 0.5*2.54)*exp(-0.5*2.54)*2 + (1-exp(-0.5*2.54))*2 });
\end{tikzpicture}
\caption{(Color online) Linearization of a surface plasmon-like waveguide dispersion relation $\tilde{\omega}(\beta)$ around a wavevector $\beta_0$ is shown. The slope of the line is equal to the group velocity $v_g$. The photon states in the text are assumed to have frequencies in the vicinity of $\omega_0$ so that the linearization is justified.}
\label{figDispersion}
\end{center}
\end{figure}
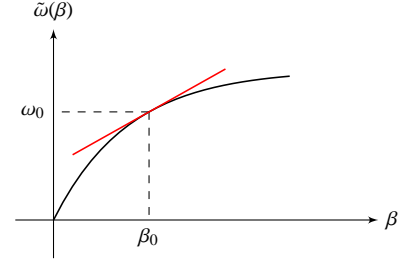

It will be useful to have $\tilde{H}$ in terms of the frequency of the photons instead of their wavevector, therefore, we linearize the waveguide dispersion around $(\beta_0,\omega_0)$ as $\tilde{\omega}(\beta) = \omega_0 + v_g (\beta-\beta_0)$ (see Fig \ref{figDispersion}). Notice that the total excitation operator 
\begin{equation}
N_E = \int_0^\infty \dif\beta \,  \tilde{a}_{\beta}^\dag \tilde{a}_{\beta} + \frac{1}{2} \sigma_z
\end{equation}
commutes with $\tilde{H}$, i.e $[\tilde{H}, N_E] = 0$. We could thus equivalently solve a system as described by 
\begin{equation}
\label{H_general}
H = \tilde{H} - \omega_0 N_E = H_0 + H_1
\end{equation}
where  
\begin{equation} 
\label{H_0}
H_0 = \int_{-\infty}^\infty \dif\beta \, v_g (\beta-\beta_0) \tilde{a}^\dag_{\beta} \tilde{a}_{\beta}
\end{equation}
\begin{equation}
H_1 = \frac{1}{2} \Omega \sigma_z + V \int_{-\infty}^\infty \dif \beta \, \left( \sigma_+ \tilde{a}_\beta  + \tilde{a}^\dag_\beta \sigma_- \right).
\end{equation}
Here $\Omega = \tilde{\Omega}-\omega_0$, and we also extended the lower limit of integration to $-\infty$ so that we can define the Fourier transform of operators in the next section. Since we will be dealing with states with wavevectors around $\beta_0$, the extension of the integration limit is well justified \cite{Blow1990,Loudon2000}.  Finally, we complete our transition from wavevectors to frequencies by defining $\omega \equiv v_g \beta$, and the operator $a_\omega \equiv  \tilde{a}_{\beta+\beta_0}/\sqrt{v_g}$, which satisfies
the commutation relation $[a_\omega, a_{\omega'}^\dag]=\delta(\omega-\omega')$. As a result of all these changes, we have
\begin{align}
\label{H0w}
H_0 &= \int_{-\infty}^\infty \dif\omega \, \omega \, a^\dag_\omega a_\omega \\
\label{H1w}
H_1 &= \frac{1}{2} \Omega \sigma_z + \frac{V}{\sqrt{v_g}} \int_{-\infty}^\infty \dif\omega \, \left( \sigma_+ a_\omega  + a^\dag_\omega \sigma_- \right).
\end{align}
Throughout the paper, the labels for photon degrees of freedom, for example $k, p$, \textit{refer to photon frequency.}

%%%%%%%%%%%%%%%%%%%%%%%%%
%%%%%%%%%%%%%%%%%%%%%%%%%
\section{Connection between the scattering theory and the input-output formalism} \label{sec.connection}
In a typical scattering experiment, various input states are prepared and sent towards a scattering region. After the scattering takes place, the outgoing states of the experiment are observed, and information about the interaction is deduced. The mathematical formulation of such a process is commonly made using the scattering matrix with elements of the form
\begin{align}
S_{p_1 p_2, k_1  k_2} = \braopket{p_1 p_2}{S}{k_1 k_2}
\end{align}
where $\ket{k_1 k_2}$ denotes the input states---here given as a two particle state with energies (frequencies) $k_1$ and $k_2$---and $\ket{p_1 p_2}$ the outgoing states. These input and output states are assumed to be free states in the interaction picture that exist long before, $t\rightarrow-\infty$, and long after, $t\rightarrow\infty$, the interaction takes place. The $S$ operator, then, is equal to the evolution operator in the interaction picture, $U_I$, from time $-\infty$ to $+\infty$,
\begin{align}
S =& \lim_{\substack{t_0\rightarrow-\infty \\ t_1\rightarrow\infty}} U_I(t_1,t_0)
= \lim_{\substack{t_0\rightarrow-\infty \\ t_1\rightarrow\infty}} \me^{\mi H_0t_1} \me^{-\mi H(t_1-t_0)}\me^{-\mi H_0t_0}
\end{align}
where $H_0$ is the non-interacting part of the Hamiltonian, and $H=H_0 + H_1$ is the total Hamiltonian.\footnote{See \cite{[{}][{ Sec. 9-e and Chap. 10.}]Taylor2006} for more information about stationary scattering theory. \cite{Cushing1990} provides a historical account of the developments related to the $S$-matrix.} In order to have a more compact notation, we will drop the limits and imply $t_0 \rightarrow - \infty$ and $t_1 \rightarrow \infty$.

An equivalent way to describe the scattering is in terms of the scattering eigenstates $\ket{k_1k_2^\pm}$ that evolve in the interaction picture from a free state either in the distant past or the distant future
\begin{align}
\ket{k_1 k_2^+} \equiv& U_I(0,t_0) \ket{k_1 k_2} = \me^{\mi H t_0} \me^{-\mi H_0 t_0}\ket{k_1 k_2}\equiv\Omega_+\ket{k_1 k_2}\\
\ket{k_1 k_2^-} \equiv& U_I(0,t_1) \ket{k_1 k_2}= \me^{\mi H t_1} \me^{-\mi H_0 t_1} \ket{k_1 k_2}\equiv\Omega_-\ket{k_1 k_2}.
\end{align}
The interaction picture time evolution operators that relate scattering and free states are called the \textit{M{\o}ller wave operators}, $\Omega_\pm$. The scattering operator can equivalently be written as $S=\Omega_-^\dag \Omega_+$.\footnote{There is also an alternative definition of the scattering operator $S'=\Omega_+ \Omega_-^\dag$ which relates the incoming and outgoing scattering eigenstates, $\ket{k^+}=S' \ket{k^-}$, such that $\braopket{p}{S}{k}=\qprod{p^-}{k^+}=\braopket{p^-}{S'}{k^-}=\braopket{p^+}{S'}{k^+}$. See \cite{Goldberger1964,Newton1982} for details.} It is also possible to write the scattering matrix elements as
\begin{align}
\braopket{p_1 p_2}{S}{k_1 k_2} = \qprod{p_1 p_2^-}{k_1 k_2^+}.
\end{align}
We should note that scattering eigenstates and the free states with the same quantum numbers have the same energies, that is $H_0 \ket{k_1 k_2} = E_{k_1 k_2}\ket{k_1 k_2}$ and $H \ket{k_1 k_2^\pm} = E_{k_1 k_2}\ket{k_1 k_2^\pm}$ \cite{Taylor2006}.

It is possible to denote the scattering matrix elements by an appropriate definition of input and output operators such that
\begin{align}
\qprod{p_1 p_2^-}{k_1 k_2^+} = \braopket{0}{\aout(p_1) \aout(p_2) \aind(k_1) \aind(k_2)}{0} \label{twophotonS}
\end{align}
where
\begin{align}
\ain(k) &\equiv\Omega_+ a_k \Omega_+^\dag = e^{\mi Ht_0} e^{-\mi H_0t_0} a_k e^{\mi H_0t_0} e^{-\mi Ht_0} \label{aindefn} \\
\aout(k) &\equiv \Omega_- a_k \Omega_-^\dag =  e^{\mi  H t_1} e^{-\mi  H_0 t_1} a_k e^{\mi  H_0 t_1} e^{-\mi  H t_1} \label{aoutdefn}
\end{align}
have the property of creating input and output scattering eigenstates
\begin{align}
\aind(k) \ket{0} &= \ket{k^+} \\
\aoutd(p) \ket{0} &= \ket{p^-}
\end{align} 
and the commutation relations
$$[\ain(k), \aind(p)] = [\aout(k), \aoutd(p)] =\delta(k-p).$$

We now relate the scattering theory, as briefly sketched above, to the input-output formalism \cite{Gardiner1985, Walls2008} of quantum optics. To do so, we start by recalling the definition of the input field operator~\cite{Gardiner1985}
\begin{equation}
\label{a_in_t}
\ain(t) =  \spi \intk a_k(t_0) e^{-\mi k (t-t_0)}
\end{equation}
where $a_k(t_0) \equiv  e^{\mi H t_0} a_k e^{-\mi H t_0}$ is an operator in the Heisenberg picture. The relationship between $\ain(t)$---which is defined in the input-output formalism---and $\ain(k)$---which is defined above in \eqref{aindefn} as a result of the scattering theory---can be determined by noting that
\begin{align}
\ain(t) &=  \spi \intk \me^{\mi Ht_0} a_k \me^{-\mi Ht_0} \me^{-\mi k (t-t_0)} \\
&=  \spi \intk \me^{\mi Ht_0} \me^{-\mi H_0 t_0} a_k \me^{\mi H_0 t_0} \me^{-\mi H t_0} \me^{-\mi k t} \\
&=   \spi \intk \ain(k) \me^{-\mi k t} \label{aintk}
\end{align}
where in the second line we used the fact that $[H_0,a_k]=-k a_k$ to convert the $a_k \me^{\mi k t_0}$ term into $\me^{-\mi H_0 t_0} a_k \me^{\mi H_0 t_0}$. As a result, $\ain(k)$ provides the spectral representation of $\ain(t)$ in the limit $t_0\rightarrow -\infty$. Similarly, the output field operator in the input-output formalism
\begin{equation}
\label{a_out_t}
\aout(t) =  \spi  \intk a_k(t_1) \me^{-\mi k (t-t_1)}
\end{equation}
is related to $\aout(k)$ in the scattering theory through
\begin{equation}
\aout(t) =  \spi \intk \aout(k) \me^{-\mi k t} \label{aouttk}
\end{equation}
 in the limit $t_1\rightarrow \infty$. We have thus established a direct connection between the input-output formalism, and the scattering theory. We should note that a different set of input and output operators were defined in \cite{Dalton1999} with an aim to make a connection to correlation functions. In \cite{Glauber1991}, a similar set of input-output operators were defined in order to relate two different quantization schemes in dielectric media. To the best of our knowledge, the explicit link we provide above between the input-output formalism and the scattering theory has not been previously published in the literature.

%%%%%%%%%%%%%%%%%%%%%%%%%
%%%%%%%%%%%%%%%%%%%%%%%%%
\section{Single-photon transport} \label{sec.onePhoton}
Now that we know the relationship between the input-output formalism and the scattering theory, let us now calculate the $S$-matrix elements $\braopket{p}{S}{k}$ between two single photon states $\ket{k}$ and $\ket{p}$. Following the standard procedure, (see Appendix \ref{appInOut}), the input-output equations appropriate for the Hamiltonian in \eqref{H_general} are
\begin{align}
\label{dN}
& \frac{\dif N}{\dif t} = - \mi \sqrt{\frac{2}{\tau}}( \sigma_+ \ain - \aind \sigma_-) - \frac{2}{\tau} N \\
\label{dsigma}
& \frac{\dif \sigma_-}{\dif t} = \mi  \sqrt{\frac{2}{\tau}} \sigma_z \ain -  \frac{1}{\tau} \sigma_-  - \mi \Omega \sigma_- \\
\label{aout}
& \aout = \ain - \mi \sqrt{\frac{2}{\tau}} \sigma_- 
\end{align}
where all operators are in the Heisenberg picture and hence they are all time-dependent. $\tau^{-1} = \pi V^2/v_g$ is proportional to the spontaneous emission rate. $N =\sigma_+ \sigma_-= ( \sigma_z + 1)/2$ describes the probability of having the atom in the excited state. 

The single-photon transport properties are described by the single photon $S$-matrix, which is related to the input and output operator by
\begin{equation}
\label{oneS}
\braopket{p}{S}{k} = \braopket{0}{\aout(p) \aind(k)}{0} = \spi \int \dif t\, \braopket{0}{\aout(t)}{k^+} \me^{\mi p t}
\end{equation}
where we used \eqref{aouttk} to write $\aout(p)$ in terms of $\aout(t)$. It is therefore sufficient to first calculate $ \braopket{0}{\aout(t)}{k^+}$ and then perform an inverse Fourier transformation to determine the single-photon $S$-matrix. In the calculations to follow in this and the next section, we will go back and forth between Fourier transforms of the operators, and we will explicitly use $t$, $t'$ to imply time dependent operators and $k_{1,2}$, $p_{1,2}$ to denote the time independent Fourier transformed pairs.

The quantity $\braopket{0}{\aout(t)}{k^+}$  can be obtained by sandwiching \eqref{dsigma} and \eqref{aout} between the two states $\bra{0}$ and $\ket{k^+}$.  We have
\begin{align}
\label{sigma0k}\begin{split}
\frac{\dif}{\dif t} \bra{0} \sigma_- \ket{k^+} =&  \mi  \sqrt{\frac{2}{\tau}} \bra{0} \sigma_z \ain \ket{k^+}-  \frac{1}{\tau} \bra{0} \sigma_-   \ket{k+} \\
&- \mi \Omega \bra{0} \sigma_- \ket{k^+} 
\end{split} \\
\label{aout0k}
\bra{0} \aout \ket{k^+}  =&\bra{0}  \ain \ket{k^+}  - \mi \sqrt{\frac{2}{\tau}} \bra{0} \sigma_- \ket{k^+}.
\end{align} 
Note that
\begin{equation}
\bra{0} \ain(t) \ket{k^+} =  \bra{0} \ain(t) \aind(k) \ket{0} =  \spi  \me^{-\mi k t} \label{ainTip}
\end{equation}
by the use of \eqref{aintk} and
\begin{equation}
\bra{0} \sigma_z \ain(t) \ket{k^+} = - \bra{0} \ain(t) \ket{k^+} \label{sigmazTip}
\end{equation}
since $\ket{0}$ has an atomic part that is in the ground state. Using \eqref{ainTip}--\eqref{sigmazTip} in \eqref{sigma0k}--\eqref{aout0k} results in a first order ordinary differential equation. By solving it, we get
\begin{align}
\bra{0} \sigma_- \ket{k^+}  &=  \spi \me^{-\mi k t} \frac{ \sqrt{2/\tau}} { ( k- \Omega ) + \mi /\tau} \label{sigk+}\\
\bra{0} \aout \ket{k^+}  &=     \spi \me^{-\mi k t} \frac{ ( k - \Omega ) - \mi/\tau} { ( k - \Omega ) + \mi/\tau }. \label{aoutkt}
\end{align}
After Fourier transforming \eqref{aoutkt}, we obtain the single-photon $S$-matrix 
\begin{equation}
\bra{p} S \ket{k} =  t_k \delta(k-p) \label{psk}
\end{equation}
where 
\begin{equation}
t_k \equiv  \frac{ ( k - \Omega ) - \mi/\tau} { ( k - \Omega ) + \mi/\tau } \label{tksk}
\end{equation}
is the single-photon transmission coefficient. For subsequent calculations, we also define 
\begin{equation}
s_k \equiv \frac{ \sqrt{2/\tau}} { ( k- \Omega ) + \mi/\tau}
\end{equation}
that measures the excitation of the atom by the single-photon wave when normalized against the incident wave amplitude. $t_k$ and $s_k$ are related by
 \begin{equation}
 \label{tfroms}
 t_k = 1 -\mi \sqrt{\frac{2}{\tau}} s_k.
 \end{equation}
These results for single-photon transport agree with \cite{Shen2007a, Shen2005}, where the scattering wavefunction was directly calculated through a real space formalism. 

The crucial step in the derivation above is \eqref{sigmazTip} which takes advantage of the single-excitation nature of the input state. Formally, the same result can also be obtained by approximately setting $\sigma_z = -1$ in \eqref{dsigma}, and thus linearizing the operator equation. Such a procedure has been commonly adopted in many quantum optics calculations \cite{Thompson1992,Domokos2002,Waks2006}. Typically, such an approximation is justified by assuming a so-called  \emph{weak excitation limit}, where the atom is assumed to be mostly in the ground state. Physically, in the case of  single-photon transport, the weak excitation limit is valid, when a single-photon pulse has a duration that is much longer than the spontaneous lifetime of the atom. However, we emphasize that the weak-excitation limit is not always valid in general even for a single-photon pulse. It has been shown that for the Hamiltonian in \eqref{H_general}, a single photon pulse with a duration comparable to the spontaneous emission lifetime can in fact \emph{completely} invert an atom \cite{Rephaeli2010}.

The formalism here removes the need for the assumption of weak-excitation limit when calculating single-photon properties. In fact, we can directly calculate the excitation probability $\braopket{k^+}{N}{k^+}$ for the scattering eigenstate $\ket{k^+}$. $N = \sigma_+ \sigma_-$ and using \eqref{sigk+} we have
\begin{align}
\bra{k^+} N \ket{k^+} &= \braopket{k^+}{\sigma_+ \sigma_-}{k^+} = \bra{k^+}\sigma_+ \ket{0} \bra{0} \sigma_- \ket{k^+} \\&= \frac{1}{2 \pi} | s_k |^2 = \frac{1}{2 \pi} \frac{ 2/\tau }  { (k - \Omega)^2 + (1/\tau)^2  }.
\end{align}
Here, we again have taken advantage of the fact that $\ket{k^+}$ is a single-excitation state whereas $\sigma_+$ acting on any state except $\ket{0}$ would result in a multi-excitation state leading to a zero overlap with $\bra{k^+}$. More generally, we have
\begin{align}
\bra{k^+} \sigma_z(t) \ket{p^+}  &= \bra{k^+} (2\sigma_+ \sigma_-  - 1) \ket{p^+}   \\
&= 2 \bra{k^+} \sigma_+ \ket{0} \bra{0} \sigma_- \ket{p^+} - \delta(k-p)    \\
&= \frac{1}{ \pi} \me^{-\mi (p-k)t} s_k^* s_p  - \delta(k-p) \label{kszp}
\end{align}
which will be useful when deriving the two-photon $S$-matrix.

%%%%%%%%%%%%%%%%%%%%%%%%%
%%%%%%%%%%%%%%%%%%%%%%%%%
\section{Two-photon transport} \label{sec.twoPhoton}
Our aim in this section is to calculate the two-photon $S$-matrix based on the results we obtained for the single photon case. We first introduced the two photon $S$-matrix element in \eqref{twophotonS}. We will begin by inserting an identity operator in between $\aout(p_1)$ and $\aout(p_2)$
\begin{align}
&\bra{0} \aout(p_1) \aout(p_2) \aind(k_1)  \aind(k_2) \ket{0} \\
&=\bra{0} \aout(p_1) \left(\int \dif k\, \ket{k^+}\bra{k^+}\right) \aout(p_2) \aind(k_1)  \aind(k_2) \ket{0}
\shortintertext{and use the Fourier transform of \eqref{aoutkt} to simplify the expression as}
&= t_{p_1} \bra{p_1^+}  \aout(p_2) \aind(k_1)  \aind(k_2) \ket{0}. \\
\shortintertext{Using the Fourier transform of \eqref{aout} we get}
&= t_{p_1}  \bra{p_1^+}  \left( \ain(p_2) - \mi \sqrt{\frac{2}{\tau}} \sigma_-(p_2)  \right) \aind(k_1)  \aind(k_2) \ket{0} \\
&=t_{p_1} \delta(p_1-k_1) \delta(p_2-k_2) + t_{p_1} \delta(p_1-k_2) \delta(p_2-k_1) \\
&\quad -\mi \sqrt{\frac{2}{\tau}}  t_{p_1} \bra{p_1^+}   \sigma_-(p_2) \ket{k_1 k_2^+}
\end{align}
where we used the orthogonality of the scattering eigenstates. Thus, to determine the two-photon $S$-matrix, we will need to calculate $\bra{p_1^+}   \sigma_-(t) \ket{k_1 k_2^+}$ and take its Fourier transform.

Using \eqref{dsigma}, we obtain the differential equation that describes $\bra{p_1^+}   \sigma_-(t) \ket{k_1 k_2^+}$
\begin{align}
\label{dsz2}\begin{split}
&\frac{\dif}{\dif t} \bra{p_1^+}   \sigma_-(t) \ket{k_1 k_2^+} \\
&= \mi \sqrt{ \frac{2}{\tau} }   \bra{p_1^+}   \sigma_z(t) \ain(t) \ket{k_1 k_2^+} - \left(\frac{1}{\tau} + \mi \Omega\right) \bra{p_1^+}   \sigma_-(t) \ket{k_1 k_2^+}.
\end{split}
\end{align}
If we can simplify the part that depends on $\sigma_z \ain$, we can then solve the differential equation. Since $\ain$ is an annihilation operator for scattering states, by using \eqref{aintk} we can write
\begin{align}
&\bra{p_1^+}   \sigma_z(t) \ain(t) \ket{k_1 k_2^+}   \\
&= \spi \left[ \bra{p_1^+}   \sigma_z(t)  \ket{k_2^+}  \me^{-\mi k_1 t} + \bra{p_1^+}   \sigma_z(t)  \ket{k_1^+}  \me^{-\mi k_2 t} \right]    \\
\shortintertext{and then using \eqref{kszp} results in}
&= \spi  \frac{1}{ \pi} \me^{-\mi (k_1+k_2-p_1)t} s_{p_1}^*(s_{k_1} + s_{k_2}) \\
&\quad- \spi \delta(k_2-p_1) \me^{- \mi k_1 t} - \spi \delta(k_1-p_1) \me^{-\mi k_2 t} 
\end{align}
which is what we were after. We can now solve the first order ordinary differential equation \eqref{dsz2} in a way very similar to the derivation that led to \eqref{psk}. After some algebra and rearrangement we get
\begin{align}
&  \bra{p_1^+}  \sigma_-(t) \ket{k_1 k_2^+}    \\
&= - \spi \frac{1}{ \pi} s_{k_1+k_2-p_1} s_{p_1}^*(s_{k_1} + s_{k_2}) \me^{-\mi (k_1+k_2-p_1)t}     \\
&\quad + \spi \delta(k_2-p_1) s_{k_1}  \me^{- \mi k_1 t} + \spi \delta(k_1-p_1) s_{k_2}   \me^{- \mi k_2 t}.
\end{align}
Taking the Fourier transform of the expression above gives us
\begin{align}
&  \bra{p_1^+}  \sigma_-(p_2) \ket{k_1 k_2^+}    \\
&=  -  \frac{1}{ \pi} \delta(k_1+k_2-p_1-p_2) s_{p_2} s_{p_1}^*(s_{k_1} + s_{k_2})    \\
&\quad + s_{k_1} \delta(k_2-p_1)   \delta(k_1-p_2) +  s_{k_2} \delta(k_1-p_1)   \delta(k_2-p_2).
\end{align}
Lastly, using the relation $t_{p_1} s_{p_1}^* = s_{p_1}$, we obtain
\begin{align}
& \bra{0} \aout(p_1) \aout(p_2) \aind(k_1)  \aind(k_2) \ket{0} \\
\label{S2} \begin{split}
& = t_{k_1} t_{k_2}  [\delta(k_2-p_1)   \delta(k_1-p_2) + \delta(k_1-p_1)   \delta(k_2-p_2)] \\
& \quad + \mi \frac{1}{\pi} \sqrt{\frac{2}{\tau}} \delta(k_1+k_2-p_1-p_2) s_{p_1} s_{p_2}(s_{k_1} + s_{k_2}).
\end{split}
\end{align}
This final result agrees with previous calculations using advanced techniques such as the Bethe ansatz\footnote{Equations (118)-(119) in \cite{Shen2007} and equation \eqref{S2} in this paper are the same with the following notational correspondence: $\Gamma = 2/\tau$, $\Delta_1=(k_1-k_2)/2$, $\Delta_2=(p_1-p_2)/2$, $E_1=k_1+k_2$, $E_2=p_1+p_2$.} in real space \cite{Shen2007}, the algebraic Bethe ansatz \cite{Yudson2008}, and the LSZ formalism in quantum field theory \cite{Shi2009,Shi2010}. The derivation here, however, is perhaps more elementary, and thus may serve to make such results more accessible. In addition, the results relate the presence of the background fluorescence, to the excitation of the atoms.

%%%%%%%%%%%%%%%%%%%%%%%%%
%%%%%%%%%%%%%%%%%%%%%%%%%
\section{Coherent State Computation} \label{sec.coherent}
A traditional use of the input-output formalism is to calculate the correlation function when the input is in a coherent state. Here we briefly outline such a calculation for our system in order to contrast it with the single and two-photon calculations of the previous two sections. For this purpose, we consider a coherent input state $\ket{\alpha_k}$, such that 
\begin{equation}
\ain(t) \ket{\alpha_k^+} = \alpha  \me^{-\mi k t} \ket{\alpha_k^+}
\end{equation}
and calculate, as an example, the $G^{(1)}$ correlation function
\begin{equation}
G^{(1)}(t', t) = \frac{ \bra{ \alpha_k^+} \aoutd(t') \aout(t) \ket{ \alpha_k^+}}{\qprod{ \alpha_k^+}{\alpha_k^+}}.
\end{equation} 
Using \eqref{aout}, we have
\begin{align}
\label{g1} \begin{split}
G^{(1)}(t, t') =& |\alpha|^2 \me^{-\mi k (t-t')} +  \mi \alpha \me^{-\mi k t} \sqrt{\frac{2}{\tau}} \expect{ \sigma_+(t') }  \\
&- \mi  \alpha^* \me^{\mi k t'} \sqrt{\frac{2}{\tau}} \expect{\sigma_-(t)}  + \frac{2}{\tau}  \expect{\sigma_+(t') \sigma_-(t)  }
\end{split}
\end{align}
where for any operator $O$, $\expect{O} \equiv  \bra{ \alpha_k^+} O  \ket{ \alpha_k^+}$.

Each of the expectation values in \eqref{g1} can be calculated using the input-output formalism. Taking the expectation values in \eqref{dN} and \eqref{dsigma} results in
\begin{align}
\frac{\dif}{\dif t} \expect{\sigma_z(t)} =& -\mi 2 \sqrt{\frac{2}{\tau}} \Big( \alpha \me^{-\mi k t} \expect{\sigma_+(t)}  
- \alpha^* \me^{\mi k t} \expect{\sigma_-(t)} \Big) \\
& -\frac{2}{\tau} \expect{\sigma_z(t)+1}     \\
\frac{\dif}{\dif t} \expect{\sigma_-(t)} =&  \left(-\mi \Omega - \frac{1}{\tau} \right) \expect{\sigma_-(t)} + \mi \alpha \me^{-\mi k t} \sqrt{\frac{2}{\tau}} \expect{\sigma_z(t)} \\
\frac{\dif}{\dif t} \expect{\sigma_+(t)} =& \left(  \mi \Omega - \frac{1}{\tau} \right) \expect{\sigma_+(t)} - \mi \alpha^* \me^{\mi k t} \sqrt{\frac{2}{\tau}} \expect{\sigma_z(t)}.
\end{align}
Directly solving the equations above provides the values of  $ \expect{ \sigma_+(t') }$ and $\expect{\sigma_-(t)}$ in \eqref{g1}, while the $\expect{\sigma_+(t') \sigma_-(t)  }$ term can be computed using the quantum regression theorem. These calculations can be found in standard textbooks \cite{Scully2008,Walls2008}, in sections related to the properties of resonance fluorescence, and we will not repeat them here. Instead, based on the outline above, we make a few observations about the coherent state computations, as commonly done, and  the one and two-photon computations as carried out in this paper.

1. The input-output formalism provides a set of \emph{nonlinear} operator equations. Therefore, all computations, by necessity, involve the conversion of such operator equations into ordinary differential equations for various operator matrix elements. While the coherent state computations typically involve taking expectation values in terms of the input states, the one and two-photon computations involve matrix elements that have different photon numbers. 

2. It is certainly reasonable to expect that the one or two-photon $S$ matrices can be obtained by analyzing various correlation functions for a weak coherent state input. Indeed, the connection between the two-photon out wavefunction, and the $g^{(2)}$ correlation function, has been pointed out in \cite{Shen2007} and it is likely that stronger connections exist. This will be carried out in future work. However, if the aim is to determine the $S$-matrix in the few-photon Fock state Hilbert space, the computation as discussed here should be far more direct. 

3. We emphasize that the few-photon computations yield the $S$-matrix in the few-photon Hilbert space, and thus provide a \emph{complete} description of all physical processes in the few-photon Fock state Hilbert space. In contrast, computing $G^{(1)}$ or $G^{(2)}$ correlation functions alone do not completely specify the out state for a given incident coherent state in general. Certainly, in the majority of quantum optics experiments at present, one probes a quantum system with a coherent input state, and obtains information about the system by measuring different correlation functions. The coherent state computations, as briefly reproduced above, are adequate to describe these experiments. However, these quantum systems are beginning to be considered as prospective devices which will eventually process quantum states \cite{Gough2009,Kerckhoff2010}. In such an engineering context, one ultimately has to be able to completely specify the output quantum states. It is in this respect that we hope the few-photon transport computations will prove to be valuable for future engineering applications.

%%%%%%%%%%%%%%%%%%%%%%%%%
%%%%%%%%%%%%%%%%%%%%%%%%%
\section{Conclusion} \label{sec.conclusion}
In this paper, we extend the input-output formalism of quantum optics to analyze one and two photon scattering in waveguides with a two-level atom inside. We develop the relationship between the input-output operators and the scattering theory which in turn enables us to analytically calculate the photon scattering matrix elements with minimum amount of algebra. We also contrast our calculations for few-photon Fock state transport with the conventional application of input-output formalism for coherent-state transport. This work helps us go beyond the correlation function analysis in input-output formalism, and leads to exact solutions for the scattering matrix elements.

%%%%%%%%%%%%%%%%%%%%%%%%%
%%%%%%%%%%%%%%%%%%%%%%%%%
\begin{acknowledgments}
This work is supported by the David \textit{\&} Lucile Packard Foundation. 
\end{acknowledgments}

%%%%%%%%%%%%%%%%%%%%%%%%%
%%%%%%%%%%%%%%%%%%%%%%%%%
\appendix

\section{Two Mode Model} \label{Appendix.TwoWay}
In this section we will write the Hamiltonian for the case when photons are allowed to propagate in both directions within the waveguide. We will refer to this case as the two mode model. After introducing the Hamiltonian, we will use the results of Sections \ref{sec.onePhoton} and \ref{sec.twoPhoton} to calculate one and two photon reflection and transmission coefficients for right to left moving fields. 

When photons propagate to the $r$ight and to the $\ell$eft, we will need to add extra terms to the Hamiltonian. We begin as we did in Section \ref{sec.hamiltonian} and write
\begin{align}
\tilde{H}_{0} = \int_0^\infty \dif \beta\, \wR(\beta) \, \aRd{\beta} \aR{\beta} 
	+ \int_{-\infty}^0 \dif \beta\, \wL(\beta) \, \aLd{\beta} \aL{\beta}
\end{align}
for the waveguide part of the Hamiltonian. The dispersion relation for the left moving modes $\wL(\beta)$ is the mirror image of the one for the right moving modes. We linearize the left and right branches of the dispersion relationship at $\beta=\pm \beta_0$ to get $\wR \approx \omega_0 +v_g (\beta-\beta_0)$ and $\wL \approx \omega_0 - v_g (\beta+\beta_0)$. Following linearization, we extend the limits of integration to $\pm\infty$, make a change of variables $\beta \mapsto \beta\mp\beta_0$ for the right and left waveguides respectively and define $\omega=v_g \beta$, $\aR{\omega} \equiv \aR{\beta+\beta_0}/\sqrt{v_g}$, $\aL{\omega} \equiv \aL{\beta -\beta_0}/\sqrt{v_g}$ to get
\begin{align}
H_{0} = \int_{-\infty}^{\infty} \dif \omega\, \omega \left( \aRd{\omega} \aR{\omega} - \aLd{\omega} \aL{\omega} \right). \label{H20}
\end{align}
The interaction part of the Hamiltonian is given by
\begin{align}
H_{1} = \frac{1}{2} \Omega \sigma_z + \frac{V}{\sqrt{v_g}} \int_{-\infty}^\infty \dif \omega \, \left[ \sigma_+ (\aR{\omega} + \aL{\omega}) + (\aRd{\omega} + \aLd{\omega}) \sigma_- \right]. \quad \label{H21}
\end{align}
Since the total excitation operator
$$
N_{E} = \int_0^\infty \dif\beta \,  \aRd{\beta} \aR{\beta} + \int_{-\infty}^0 \dif\beta \,  \aLd{\beta} \aL{\beta} + \frac{1}{2} \sigma_z
$$
commutes with the Hamiltonian, we subtracted the term $\omega_0 N_{E}$ from the Hamiltonian and set $\Omega = \tilde{\Omega} - \omega_0$ in the derivation, mimicking the steps in Section \ref{sec.hamiltonian}.

Now that we have the Hamiltonian, we can write down the Heisenberg equations of motion and define the input-output operators for the fields as illustrated in detail for a chiral model in Appendix \ref{appInOut}. The equations for the annihilation operators are
\begin{align}
\frac{\dif \aR{\omega}(t)}{\dif t}=-\mi[\aR{\omega},H]=-\mi \omega \aR{\omega} - \mi \Vt \sigma_- \\
\frac{\dif \aL{\omega}(t)}{\dif t}=-\mi[\aL{\omega},H]=+\mi \omega \aL{\omega} - \mi \Vt \sigma_-
\end{align}
where $\Vt = V/\sqrt{v_g}$. The definitions for the input and output operators for right going fields are the same as in Appendix \ref{appInOut} and we get
\begin{align}
\aRout(t) = \aRin(t) - \mi \sqrt{\frac{2}{\tau}}\sigma_-(t)
\end{align}
where $\tau$ is defined in \eqref{taudefn}. Left going modes have a group velocity which is negative that of the right going modes and that leads to a negative sign in \eqref{H20}. As a result, starting from the the definition of the input and output operators in \eqref{aindefn}--\eqref{aoutdefn}, the input and output operators for left going modes have the form
\begin{align}
\aLout(t) &= \spi \int \dif \omega \, \aL{\omega}(t_1) \me^{\mi \omega (t-t_1)}\\
\aLin(t) &= \spi \int \dif \omega \, \aL{\omega}(t_0) \me^{\mi \omega (t-t_0)}\\
\aLout(t) &= \aLin(t) -  \mi \sqrt{\frac{2}{\tau}}\sigma_-(t)
\end{align}
where we note the change of sign in the frequency variable. Using these results we can show that
$$
\frac{\dif \sigma_-}{\dif t} 	= \mi \sqrt{\frac{2}{\tau}} \sigma_z \aRin + \mi \sqrt{\frac{2}{\tau}} \sigma_z \aLin -  \frac{2}{\tau} \sigma_-  - \mi \Omega \sigma_-
$$
which is in a form similar to those that we get in temporal coupled mode theory \cite{[{}][{ Sec 7.4.}]Haus1984}.

We now have all the tools to solve for the scattering that takes place in the two mode model. Let us define even and odd combinations of the operators for the right and left propagating modes as
\begin{align}
\label{evenodd}
a_\omega = \frac{ \aR{\omega} + \aL{-\omega} }{\sqrt{2}} \: \text{(even)}  \qquad 
\mathring{a}_\omega = \frac{ \aR{\omega} - \aL{-\omega} }{\sqrt{2}} \: \text{(odd).}
\end{align}
Using these definitions in \eqref{H20}--\eqref{H21} we can show
\begin{align}
H_{0}&=\int \dif \omega \, a_\omega^\dag a_\omega + \mathring{a}_\omega^\dag \mathring{a}_\omega \equiv H_{\text{e},0} + H_{\text{o},0}\\
H_{1}& = \frac{1}{2} \Omega \sigma_z + \frac{\sqrt{2}V}{\sqrt{v_g}}\int \dif \omega \, \left( \sigma_+ a_\omega + a_\omega^\dag \sigma_- \right) \equiv H_{\text{e},1} 
\end{align}
where we see that the interaction part of the Hamiltonian depends only on the even combination of modes. In Sections \ref{sec.onePhoton} and \ref{sec.twoPhoton} we solved for $H=H_{\text{e},0}+H_{\text{e},1} $ for a rescaled value of $V$. The odd part $H_{\text{o},0}$ is interaction free and hence is also solved. From \eqref{evenodd} we get 
\begin{align}
\label{decomposition}
\begin{split}
\aRinout(\omega) &= \frac{\ainout(\omega)+\azinout(\omega)}{\sqrt{2}} \\
\aLout(\omega) &= \frac{\aout(-\omega)-\azout(-\omega)}{\sqrt{2}}
\end{split}
\end{align}
where we wrote the Fourier transforms of two-mode input-output operators in terms of the combinations of even and odd fields. 

The get the one photon reflection probability, we look at the scattering matrix element which corresponds to a right propagating input photon and a left propagating output photon
\begin{align}
&\braopket{0}{\aLout(p) \aRind(k)}{0} \\
&= \frac{1}{2} \braopket{0}{[\aout(-p)-\azout(-p)][\aind(k)+\azind(k)]}{0}\\
&=\frac{1}{2} \braopket{0}{\aout(-p)\aind(k)}{0} - \frac{1}{2} \braopket{0}{\azout(-p)\azind(k)}{0} \\
&=\frac{1}{2}(t_k-1) \delta(p+k) \equiv \bar{r}_k\delta(p+k).
\end{align}
Here we used \eqref{decomposition} and \eqref{psk} to get the one photon reflection coefficient $\bar{r}_k$. Similarly, the one photon transmission coefficient $\bar{t}_k$ is given by
\begin{align}
&\braopket{0}{\aRout(p) \aRind(k)}{0} =\frac{1}{2}(t_k+1) \delta(p-k) \equiv \bar{t}_k\delta(p-k).
\end{align}

Two photon calculations require adding another input-output pair. For instance, the scattering matrix element associated with one photon scattering to the right, another to the left when two photons initially propagate to the right is given by
\begin{align}
&\braopket{0}{\aRout(p_1) \aLout(p_2) \aRind(k_1) \aRind(k_2) }{0} \\
&=\frac{1}{4}\Big[ \braopket{0}{\aout(p_1) \aout(-p_2) \aind(k_1) \aind(k_2)}{0} \\
& \qquad - \braopket{0}{\aout(p_1) \azout(-p_2) \aind(k_1) \azind(k_2)}{0} \\
& \qquad - \braopket{0}{\aout(p_1) \azout(-p_2) \azind(k_1) \aind(k_2)}{0} \\
& \qquad + \braopket{0}{\azout(p_1) \aout(-p_2) \aind(k_1) \azind(k_2)}{0} \\
& \qquad + \braopket{0}{\azout(p_1) \aout(-p_2) \azind(k_1) \aind(k_2)}{0} \\
& \qquad - \braopket{0}{\azout(p_1) \azout(-p_2) \azind(k_1) \azind(k_2)}{0} \Big]\\
&=\bar{t}_{k_1} \bar{r}_{k_2} \delta(k_1-p_1)\delta(k_2+p_2) + \bar{r}_{k_1} \bar{t}_{k_2} \delta(k_1+p_2)\delta(k_2-p_1)\\ &\quad + \frac{1}{4} B \delta(k_1+k_2-p_1+p_2)
\end{align}
where from \eqref{S2}
$$B=\mi \frac{1}{\pi} \sqrt{\frac{2}{\tau'}} s_{p_1} s_{-p_2}(s_{k_1} + s_{k_2}).$$
We note that $\tau' = \tau/2$ due to an extra factor of $\sqrt{2}$ before $V$ in the definition of $H_{1}$. These results agree with equations (52) and (130) in \cite{Shen2007}.

%%%%%%%%%%%%%%%%%%%%%%%%%
%%%%%%%%%%%%%%%%%%%%%%%%%

\section{Hamiltonian in the continuum limit} \label{Appendix.Continuum}
This section will summarize the steps taken to obtain the continuum form of the Hamiltonian from its discrete version. We will follow the approach in \cite{*[{}] [{ Sec. 6.2.}] Loudon2000,Blow1990}.

The discrete variables are assumed to be for those in a one dimensional cavity of length $L$. The mode spacing in the cavity is given by $\Delta \beta = 2\pi/L$. In this 1D cavity, the free space electromagnetic Hamiltonian, $H_0$, is given by
\begin{align}
H_0= \sum_\beta  \omega_\beta \, \hat{a}_\beta^\dag \hat{a}_\beta
\end{align}
with the commutator relationship $[\hat{a}_\beta,\hat{a}_{\beta'}^\dag]=\delta_{\beta,\beta'}$. Now, we will convert the sum into an integral by the equivalence $\left(\Delta \beta \sum_\beta\right) \rightarrow \left( \int \dif \beta\right)$ to get
\begin{align}
H_0=\frac{L}{2\pi} \int \dif\beta\, \omega_\beta \, \hat{a}_\beta^\dag \hat{a}_\beta .
\end{align}
The continuous mode operator $\tilde{a}_\beta$ is related to the discrete mode $\hat{a}_\beta$ by
\begin{align}
\tilde{a}_\beta=\sqrt{\frac{L}{2\pi}} \hat{a}_\beta \quad \text{which results in}\quad
H_0=\int \dif\beta\, \omega_\beta \, \tilde{a}_\beta^\dag \tilde{a}_\beta .
\end{align}
The commutator relationship $[\tilde{a}_\beta, \tilde{a}_{\beta'}^\dag]=\frac{L}{2\pi} \delta_{\beta,\beta'}$ in the limit $L \rightarrow \infty$ becomes
\begin{align}
[\tilde{a}_\beta, \tilde{a}_{\beta'}^\dag]=\delta(\beta-\beta') .
\end{align}
To see this result, define $f(\beta) = \frac{L}{2\pi} \delta_{\beta,0}$. Integrating $f(\beta)$ will give
\begin{align}
\int \dif\beta\, f(\beta) \rightarrow \frac{2 \pi}{L} \sum_\beta f(\beta) = \frac{2\pi}{L} \frac{L}{2\pi} = 1.
\end{align}
As a result, the correct Hamiltonian in the continuum limit is
\begin{align}
H_0=\int \dif\beta\, \omega(\beta) \, \tilde{a}_\beta^\dag \tilde{a}_\beta 
\end{align}
with $[\tilde{a}_{\beta}, \tilde{a}_{\beta'}^\dag]=\delta(\beta-\beta')$. It is then easy to show that
\begin{align}
\mathbf{1} = \int \dif \beta\, \ket{\beta} \bra{\beta} 
\end{align}
where $\ket{\beta}=\tilde{a}_\beta^\dag \ket{0}$, since
\begin{align}
\bra{\gamma}\int \dif \beta\, \ket{\beta}\qprod{\beta}{\zeta} = \int \dif \beta\, \delta(\gamma-\beta) \delta(\beta-\zeta) = \delta(\gamma-\zeta).
\end{align}

In the discreet case 
\begin{align}
H_1=\frac{1}{2} \tilde{\Omega} \sigma_z + \frac{V'}{\sqrt{L}} \sum_\beta (\sigma_+ \hat{a}_\beta + \hat{a}_\beta^\dag \sigma_-)
\end{align}
where $V'$ is the physical coupling constant. The factor $L^{-1/2}$ arises because the photon as created by $\hat{a}_\beta^\dag$ has a normalization constant $L^{-1/2}$. In the continuum case we get 
\begin{align}
H_1 =& \frac{1}{2} \tilde{\Omega} \sigma_z + \frac{V'}{\sqrt{L}} \frac{L}{2 \pi} \sqrt{\frac{2 \pi}{L}} \int \dif \beta\, (\sigma_+ \tilde{a}_\beta + \tilde{a}_\beta^\dag \sigma_-)\\
 =& \frac{1}{2} \tilde{\Omega} \sigma_z + \frac{V'}{\sqrt{2 \pi}} \int \dif\beta\, (\sigma_+ \tilde{a}_\beta + \tilde{a}_\beta^\dag \sigma_- ).
\end{align}
Thus, the coupling constants in the discrete ($V'$) and the continuum ($V$) cases differ by a factor of $(2 \pi)^{-1/2}$.

%%%%%%%%%%%%%%%%%%%%%%%%%
%%%%%%%%%%%%%%%%%%%%%%%%%

\section{Derivation of the input-output formalism} \label{appInOut}
Here we provide a derivation of the input-output equations \eqref{dN}--\eqref{aout}. This derivation closely follows \cite{Gardiner1985, Walls2008}.  Based on the Hamiltonian \eqref{H0w}--\eqref{H1w}, and the definition $\Vt \equiv V/\sqrt{v_g}$, the Heisenberg equations of motion for the operators are
\begin{align}
\label{dak}
\mi \frac{\dif a_k}{\dif t} &= k a_k + \Vt \sigma_-\\
\label{dsigmam}
\mi \frac{\dif \sigma_-}{\dif t} &= \Omega \sigma_- - \Vt \intk \sigma_z a_k\\
\label{dsigmaz}
\mi \frac{\dif \sigma_z}{\dif t} &=  2 \Vt \intk ( -a_k^\dag \sigma_-  +   \sigma_+ a_k).\\
\end{align}
After multiplying \eqref{dak} by the integration factor $\exp(\mi k t)$, we integrate it from an initial time $t_0 < t$ to get
\begin{equation}
\label{akt}
a_k(t) = a_k(t_0) \me^{-\mi k (t-t_0)} - \mi \Vt \int_{t_0}^{t} dt'  \sigma_-(t') \me^{-\mi k (t-t')}.
\end{equation}
We define the input operator as 
\begin{equation}
\ain(t) = \frac{1}{\sqrt{2\pi}} \intk a_k(t_0) \me^{-\mi k (t-t_0)}
\end{equation}
which satisfies the commutation relation $$[\ain(t), \aind(t')] = \delta(t-t').$$ We further introduce a field operator
\begin{equation}
\Phi(t) = \frac{1}{\sqrt{2\pi}} \intk a_k(t)
\end{equation}
and integrate \eqref{akt} with respect to $k$ to get
\begin{align}
\label{phi_in} \begin{split}
\Phi(t) &= \ain(t) -\mi \frac{\Vt}{2} \sqrt{2 \pi} \sigma_-(t)= \ain(t) -\mi \sqrt{\frac{1}{2 \tau}} \sigma_-(t).\quad
\end{split}
\end{align}
Here, notice that we integrate over half the delta-function \cite{Gardiner1985} which results in a factor of $1/2$ and $\tau$ is defined as 
\begin{equation}
\frac{1}{\tau} \equiv \pi \Vt^2. \label{taudefn}
\end{equation}
Furthermore, plugging \eqref{phi_in} into \eqref{dsigmam} and \eqref{dsigmaz}, results in
\begin{align}
& \frac{\dif \sigma_-}{\dif t} 	= 	\mi \sqrt{\frac{2}{\tau}} \sigma_z \ain -  \frac{1}{\tau} \sigma_-  - \mi \Omega \sigma_- \label{app.sigmam}\\
& \frac{\dif N}{\dif t} 		= - 	\mi \sqrt{\frac{2}{\tau}}( \sigma_+ \ain - \aind \sigma_-) - \frac{2}{\tau} N.
\end{align}
Here $N =  (\sigma_z + 1)/2 $. Thus the spontaneous emission rate is ${2}/{\tau}$. We could have also directly calculated ${\dif N}/{\dif t}$ from ${\dif \sigma_-}/{\dif t}$, since $N = \sigma_+ \sigma_-$.

Similarly, we integrate \eqref{dak}  up to a final time $t_1 > t$, and define an output operator
\begin{equation}
\aout(t) = \spi \intk a_k(t_1) \me^{-\mi k (t-t_1)}
\end{equation}
which results in 
\begin{equation}
\label{phi_out}
\Phi(t) = \aout(t) + \mi \sqrt{\frac{1}{2 \tau}} \sigma_-(t) .
\end{equation}
Combining \eqref{phi_in} and \eqref{phi_out}, we finally obtain
\begin{equation}
\aout(t) = \ain(t) - \mi \sqrt{\frac{2}{\tau}} \sigma_-(t).
\end{equation}



\begin{thebibliography}{38}%
\makeatletter
\providecommand \@ifxundefined [1]{%
 \@ifx{#1\undefined}
}%
\providecommand \@ifnum [1]{%
 \ifnum #1\expandafter \@firstoftwo
 \else \expandafter \@secondoftwo
 \fi
}%
\providecommand \@ifx [1]{%
 \ifx #1\expandafter \@firstoftwo
 \else \expandafter \@secondoftwo
 \fi
}%
\providecommand \natexlab [1]{#1}%
\providecommand \enquote  [1]{``#1''}%
\providecommand \bibnamefont  [1]{#1}%
\providecommand \bibfnamefont [1]{#1}%
\providecommand \citenamefont [1]{#1}%
\providecommand \href@noop [0]{\@secondoftwo}%
\providecommand \href [0]{\begingroup \@sanitize@url \@href}%
\providecommand \@href[1]{\@@startlink{#1}\@@href}%
\providecommand \@@href[1]{\endgroup#1\@@endlink}%
\providecommand \@sanitize@url [0]{\catcode `\\12\catcode `\$12\catcode
  `\&12\catcode `\#12\catcode `\^12\catcode `\_12\catcode `\%12\relax}%
\providecommand \@@startlink[1]{}%
\providecommand \@@endlink[0]{}%
\providecommand \url  [0]{\begingroup\@sanitize@url \@url }%
\providecommand \@url [1]{\endgroup\@href {#1}{\urlprefix }}%
\providecommand \urlprefix  [0]{URL }%
\providecommand \Eprint [0]{\href }%
\providecommand \doibase [0]{http://dx.doi.org/}%
\providecommand \selectlanguage [0]{\@gobble}%
\providecommand \bibinfo  [0]{\@secondoftwo}%
\providecommand \bibfield  [0]{\@secondoftwo}%
\providecommand \translation [1]{[#1]}%
\providecommand \BibitemOpen [0]{}%
\providecommand \bibitemStop [0]{}%
\providecommand \bibitemNoStop [0]{.\EOS\space}%
\providecommand \EOS [0]{\spacefactor3000\relax}%
\providecommand \BibitemShut  [1]{\csname bibitem#1\endcsname}%
\let\auto@bib@innerbib\@empty
%</preamble>
\bibitem [{\citenamefont {O'Brien}\ \emph {et~al.}(2009)\citenamefont
  {O'Brien}, \citenamefont {Furusawa},\ and\ \citenamefont
  {Vuckovic}}]{OBrien2009}%
  \BibitemOpen
  \bibfield  {author} {\bibinfo {author} {\bibfnamefont {J.~L.}\ \bibnamefont
  {O'Brien}}, \bibinfo {author} {\bibfnamefont {A.}~\bibnamefont {Furusawa}}, \
  and\ \bibinfo {author} {\bibfnamefont {J.}~\bibnamefont {Vuckovic}},\
  }\bibfield  {title} {\enquote {\bibinfo {title} {Photonic quantum
  technologies},}\ }\href {http://dx.doi.org/10.1038/nphoton.2009.229}
  {\bibfield  {journal} {\bibinfo  {journal} {Nat Photon}\ }\textbf {\bibinfo
  {volume} {3}},\ \bibinfo {pages} {687--695} (\bibinfo {year}
  {2009})}\BibitemShut {NoStop}%
\bibitem [{\citenamefont {Kimble}(2008)}]{Kimble2008}%
  \BibitemOpen
  \bibfield  {author} {\bibinfo {author} {\bibfnamefont {H.~J.}\ \bibnamefont
  {Kimble}},\ }\bibfield  {title} {\enquote {\bibinfo {title} {The quantum
  internet},}\ }\href {http://dx.doi.org/10.1038/nature07127} {\bibfield
  {journal} {\bibinfo  {journal} {Nature}\ }\textbf {\bibinfo {volume} {453}},\
  \bibinfo {pages} {1023--1030} (\bibinfo {year} {2008})}\BibitemShut {NoStop}%
\bibitem [{\citenamefont {Schoelkopf}\ and\ \citenamefont
  {Girvin}(2008)}]{Schoelkopf2008}%
  \BibitemOpen
  \bibfield  {author} {\bibinfo {author} {\bibfnamefont {R.~J.}\ \bibnamefont
  {Schoelkopf}}\ and\ \bibinfo {author} {\bibfnamefont {S.~M.}\ \bibnamefont
  {Girvin}},\ }\bibfield  {title} {\enquote {\bibinfo {title} {Wiring up
  quantum systems},}\ }\href {http://dx.doi.org/10.1038/451664a} {\bibfield
  {journal} {\bibinfo  {journal} {Nature}\ }\textbf {\bibinfo {volume} {451}},\
  \bibinfo {pages} {664--669} (\bibinfo {year} {2008})}\BibitemShut {NoStop}%
\bibitem [{\citenamefont {Dayan}\ \emph {et~al.}(2008)\citenamefont {Dayan},
  \citenamefont {Parkins}, \citenamefont {Aoki}, \citenamefont {Ostby},
  \citenamefont {Vahala},\ and\ \citenamefont {Kimble}}]{Dayan2008}%
  \BibitemOpen
  \bibfield  {author} {\bibinfo {author} {\bibfnamefont {B.}~\bibnamefont
  {Dayan}}, \bibinfo {author} {\bibfnamefont {A.~S.}\ \bibnamefont {Parkins}},
  \bibinfo {author} {\bibfnamefont {T.}~\bibnamefont {Aoki}}, \bibinfo {author}
  {\bibfnamefont {E.~P.}\ \bibnamefont {Ostby}}, \bibinfo {author}
  {\bibfnamefont {K.~J.}\ \bibnamefont {Vahala}}, \ and\ \bibinfo {author}
  {\bibfnamefont {H.~J.}\ \bibnamefont {Kimble}},\ }\bibfield  {title}
  {\enquote {\bibinfo {title} {A photon turnstile dynamically regulated by one
  atom},}\ }\href {\doibase 10.1126/science.1152261} {\bibfield  {journal}
  {\bibinfo  {journal} {Science}\ }\textbf {\bibinfo {volume} {319}},\ \bibinfo
  {pages} {1062--1065} (\bibinfo {year} {2008})}\BibitemShut {NoStop}%
\bibitem [{\citenamefont {Wallraff}\ \emph {et~al.}(2004)\citenamefont
  {Wallraff}, \citenamefont {Schuster}, \citenamefont {Blais}, \citenamefont
  {Frunzio}, \citenamefont {Huang}, \citenamefont {Majer}, \citenamefont
  {Kumar}, \citenamefont {Girvin},\ and\ \citenamefont
  {Schoelkopf}}]{Wallraff2004}%
  \BibitemOpen
  \bibfield  {author} {\bibinfo {author} {\bibfnamefont {A.}~\bibnamefont
  {Wallraff}}, \bibinfo {author} {\bibfnamefont {D.~I.}\ \bibnamefont
  {Schuster}}, \bibinfo {author} {\bibfnamefont {A.}~\bibnamefont {Blais}},
  \bibinfo {author} {\bibfnamefont {L.}~\bibnamefont {Frunzio}}, \bibinfo
  {author} {\bibfnamefont {R.-S.}\ \bibnamefont {Huang}}, \bibinfo {author}
  {\bibfnamefont {J.}~\bibnamefont {Majer}}, \bibinfo {author} {\bibfnamefont
  {S.}~\bibnamefont {Kumar}}, \bibinfo {author} {\bibfnamefont {S.~M.}\
  \bibnamefont {Girvin}}, \ and\ \bibinfo {author} {\bibfnamefont {R.~J.}\
  \bibnamefont {Schoelkopf}},\ }\bibfield  {title} {\enquote {\bibinfo {title}
  {Strong coupling of a single photon to a superconducting qubit using circuit
  quantum electrodynamics},}\ }\href {http://dx.doi.org/10.1038/nature02851}
  {\bibfield  {journal} {\bibinfo  {journal} {Nature}\ }\textbf {\bibinfo
  {volume} {431}},\ \bibinfo {pages} {162--167} (\bibinfo {year}
  {2004})}\BibitemShut {NoStop}%
\bibitem [{\citenamefont {Astafiev}\ \emph
  {et~al.}(2010{\natexlab{a}})\citenamefont {Astafiev}, \citenamefont
  {Zagoskin}, \citenamefont {Abdumalikov{, Jr.}}, \citenamefont {Pashkin},
  \citenamefont {Yamamoto}, \citenamefont {Inomata}, \citenamefont {Nakamura},\
  and\ \citenamefont {Tsai}}]{Astafiev2010}%
  \BibitemOpen
  \bibfield  {author} {\bibinfo {author} {\bibfnamefont {O.}~\bibnamefont
  {Astafiev}}, \bibinfo {author} {\bibfnamefont {A.~M.}\ \bibnamefont
  {Zagoskin}}, \bibinfo {author} {\bibfnamefont {A.~A.}\ \bibnamefont
  {Abdumalikov{, Jr.}}}, \bibinfo {author} {\bibfnamefont {Yu.~A.}\
  \bibnamefont {Pashkin}}, \bibinfo {author} {\bibfnamefont {T.}~\bibnamefont
  {Yamamoto}}, \bibinfo {author} {\bibfnamefont {K.}~\bibnamefont {Inomata}},
  \bibinfo {author} {\bibfnamefont {Y.}~\bibnamefont {Nakamura}}, \ and\
  \bibinfo {author} {\bibfnamefont {J.~S.}\ \bibnamefont {Tsai}},\ }\bibfield
  {title} {\enquote {\bibinfo {title} {Resonance fluorescence of a single
  artificial atom},}\ }\href {\doibase 10.1126/science.1181918} {\bibfield
  {journal} {\bibinfo  {journal} {Science}\ }\textbf {\bibinfo {volume}
  {327}},\ \bibinfo {pages} {840--843} (\bibinfo {year}
  {2010}{\natexlab{a}})}\BibitemShut {NoStop}%
\bibitem [{\citenamefont {Astafiev}\ \emph
  {et~al.}(2010{\natexlab{b}})\citenamefont {Astafiev}, \citenamefont
  {Abdumalikov}, \citenamefont {Zagoskin}, \citenamefont {Pashkin},
  \citenamefont {Nakamura},\ and\ \citenamefont {Tsai}}]{Astafiev2010a}%
  \BibitemOpen
  \bibfield  {author} {\bibinfo {author} {\bibfnamefont {O.~V.}\ \bibnamefont
  {Astafiev}}, \bibinfo {author} {\bibfnamefont {A.~A.}\ \bibnamefont
  {Abdumalikov}}, \bibinfo {author} {\bibfnamefont {A.~M.}\ \bibnamefont
  {Zagoskin}}, \bibinfo {author} {\bibfnamefont {Yu.~A.}\ \bibnamefont
  {Pashkin}}, \bibinfo {author} {\bibfnamefont {Y.}~\bibnamefont {Nakamura}}, \
  and\ \bibinfo {author} {\bibfnamefont {J.~S.}\ \bibnamefont {Tsai}},\
  }\bibfield  {title} {\enquote {\bibinfo {title} {Ultimate on-chip quantum
  amplifier},}\ }\href {\doibase 10.1103/PhysRevLett.104.183603} {\bibfield
  {journal} {\bibinfo  {journal} {Phys. Rev. Lett.}\ }\textbf {\bibinfo
  {volume} {104}},\ \bibinfo {pages} {183603} (\bibinfo {year}
  {2010}{\natexlab{b}})}\BibitemShut {NoStop}%
\bibitem [{\citenamefont {Kojima}\ \emph {et~al.}(2003)\citenamefont {Kojima},
  \citenamefont {Hofmann}, \citenamefont {Takeuchi},\ and\ \citenamefont
  {Sasaki}}]{Kojima2003}%
  \BibitemOpen
  \bibfield  {author} {\bibinfo {author} {\bibfnamefont {K.}~\bibnamefont
  {Kojima}}, \bibinfo {author} {\bibfnamefont {H.~F.}\ \bibnamefont {Hofmann}},
  \bibinfo {author} {\bibfnamefont {S.}~\bibnamefont {Takeuchi}}, \ and\
  \bibinfo {author} {\bibfnamefont {K.}~\bibnamefont {Sasaki}},\ }\bibfield
  {title} {\enquote {\bibinfo {title} {Nonlinear interaction of two photons
  with a one-dimensional atom: Spatiotemporal quantum coherence in the emitted
  field},}\ }\href {\doibase 10.1103/PhysRevA.68.013803} {\bibfield  {journal}
  {\bibinfo  {journal} {Phys. Rev. A}\ }\textbf {\bibinfo {volume} {68}},\
  \bibinfo {pages} {013803} (\bibinfo {year} {2003})}\BibitemShut {NoStop}%
\bibitem [{\citenamefont {Shen}\ and\ \citenamefont {Fan}(2005)}]{Shen2005}%
  \BibitemOpen\bibliography{quantum}
  \bibfield  {author} {\bibinfo {author} {\bibfnamefont {J.~T.}\ \bibnamefont
  {Shen}}\ and\ \bibinfo {author} {\bibfnamefont {S.}~\bibnamefont {Fan}},\
  }\bibfield  {title} {\enquote {\bibinfo {title} {Coherent photon transport
  from spontaneous emission in one-dimensional waveguides},}\ }\href {\doibase
  10.1364/OL.30.002001} {\bibfield  {journal} {\bibinfo  {journal} {Opt.
  Lett.}\ }\textbf {\bibinfo {volume} {30}},\ \bibinfo {pages} {2001--2003}
  (\bibinfo {year} {2005})}\BibitemShut {NoStop}%
\bibitem [{\citenamefont {Shen}\ and\ \citenamefont
  {Fan}(2007{\natexlab{a}})}]{Shen2007}%
  \BibitemOpen
  \bibfield  {author} {\bibinfo {author} {\bibfnamefont {J.~T.}\ \bibnamefont
  {Shen}}\ and\ \bibinfo {author} {\bibfnamefont {S.}~\bibnamefont {Fan}},\
  }\bibfield  {title} {\enquote {\bibinfo {title} {Strongly correlated
  multiparticle transport in one dimension through a quantum impurity},}\
  }\href {\doibase 10.1103/PhysRevA.76.062709} {\bibfield  {journal} {\bibinfo
  {journal} {Phys. Rev. A}\ }\textbf {\bibinfo {volume} {76}},\ \bibinfo
  {pages} {062709} (\bibinfo {year} {2007}{\natexlab{a}})}\BibitemShut
  {NoStop}%
\bibitem [{\citenamefont {Shen}\ and\ \citenamefont
  {Fan}(2007{\natexlab{b}})}]{Shen2007a}%
  \BibitemOpen
  \bibfield  {author} {\bibinfo {author} {\bibfnamefont {J.~T.}\ \bibnamefont
  {Shen}}\ and\ \bibinfo {author} {\bibfnamefont {S.}~\bibnamefont {Fan}},\
  }\bibfield  {title} {\enquote {\bibinfo {title} {Strongly correlated
  two-photon transport in a one-dimensional waveguide coupled to a two-level
  system},}\ }\href {\doibase 10.1103/PhysRevLett.98.153003} {\bibfield
  {journal} {\bibinfo  {journal} {Phys. Rev. Lett.}\ }\textbf {\bibinfo
  {volume} {98}},\ \bibinfo {pages} {153003} (\bibinfo {year}
  {2007}{\natexlab{b}})}\BibitemShut {NoStop}%
\bibitem [{\citenamefont {Chang}\ \emph {et~al.}(2007)\citenamefont {Chang},
  \citenamefont {S{\o}rensen}, \citenamefont {Demler},\ and\ \citenamefont
  {Lukin}}]{Chang2007}%
  \BibitemOpen
  \bibfield  {author} {\bibinfo {author} {\bibfnamefont {D.~E.}\ \bibnamefont
  {Chang}}, \bibinfo {author} {\bibfnamefont {A.~S.}\ \bibnamefont
  {S{\o}rensen}}, \bibinfo {author} {\bibfnamefont {E.~A.}\ \bibnamefont
  {Demler}}, \ and\ \bibinfo {author} {\bibfnamefont {M.~D.}\ \bibnamefont
  {Lukin}},\ }\bibfield  {title} {\enquote {\bibinfo {title} {A single-photon
  transistor using nanoscale surface plasmons},}\ }\href {\doibase
  10.1038/nphys708} {\bibfield  {journal} {\bibinfo  {journal} {Nature
  Physics}\ }\textbf {\bibinfo {volume} {3}},\ \bibinfo {pages} {807--812}
  (\bibinfo {year} {2007})}\BibitemShut {NoStop}%
\bibitem [{\citenamefont {Roy}(2010)}]{Roy2010}%
  \BibitemOpen
  \bibfield  {author} {\bibinfo {author} {\bibfnamefont {D.}~\bibnamefont
  {Roy}},\ }\bibfield  {title} {\enquote {\bibinfo {title} {Few-photon optical
  diode},}\ }\href {\doibase 10.1103/PhysRevB.81.155117} {\bibfield  {journal}
  {\bibinfo  {journal} {Phys. Rev. B}\ }\textbf {\bibinfo {volume} {81}},\
  \bibinfo {pages} {155117} (\bibinfo {year} {2010})}\BibitemShut {NoStop}%
\bibitem [{\citenamefont {Yudson}\ and\ \citenamefont
  {Reineker}(2008)}]{Yudson2008}%
  \BibitemOpen
  \bibfield  {author} {\bibinfo {author} {\bibfnamefont {V.~I.}\ \bibnamefont
  {Yudson}}\ and\ \bibinfo {author} {\bibfnamefont {P.}~\bibnamefont
  {Reineker}},\ }\bibfield  {title} {\enquote {\bibinfo {title} {Multiphoton
  scattering in a one-dimensional waveguide with resonant atoms},}\ }\href
  {\doibase 10.1103/PhysRevA.78.052713} {\bibfield  {journal} {\bibinfo
  {journal} {Phys. Rev. A}\ }\textbf {\bibinfo {volume} {78}},\ \bibinfo
  {pages} {052713} (\bibinfo {year} {2008})}\BibitemShut {NoStop}%
\bibitem [{\citenamefont {Witthaut}\ and\ \citenamefont
  {S{\o}rensen}(2010)}]{Witthaut2010}%
  \BibitemOpen
  \bibfield  {author} {\bibinfo {author} {\bibfnamefont {D.}~\bibnamefont
  {Witthaut}}\ and\ \bibinfo {author} {\bibfnamefont {A.~S.}\ \bibnamefont
  {S{\o}rensen}},\ }\bibfield  {title} {\enquote {\bibinfo {title} {Photon
  scattering by a three-level emitter in a one-dimensional waveguide},}\ }\href
  {http://stacks.iop.org/1367-2630/12/i=4/a=043052} {\bibfield  {journal}
  {\bibinfo  {journal} {New Journal of Physics}\ }\textbf {\bibinfo {volume}
  {12}},\ \bibinfo {pages} {043052} (\bibinfo {year} {2010})}\BibitemShut
  {NoStop}%
\bibitem [{\citenamefont {Zheng}\ \emph {et~al.}(2010)\citenamefont {Zheng},
  \citenamefont {Gauthier},\ and\ \citenamefont {Baranger}}]{Zheng2010}%
  \BibitemOpen
  \bibfield  {author} {\bibinfo {author} {\bibfnamefont {H.}~\bibnamefont
  {Zheng}}, \bibinfo {author} {\bibfnamefont {D.~J.}\ \bibnamefont {Gauthier}},
  \ and\ \bibinfo {author} {\bibfnamefont {H.~U.}\ \bibnamefont {Baranger}},\
  }\href@noop {} {\enquote {\bibinfo {title} {Waveguide {QED}: Many-body bound
  state effects on coherent and fock state scattering from a two-level
  system},}\ } (\bibinfo {year} {2010}),\ \Eprint
  {http://arxiv.org/abs/1009.5325} {arXiv:1009.5325} \BibitemShut {NoStop}%
\bibitem [{\citenamefont {Liao}\ and\ \citenamefont {Law}(2010)}]{Liao2010}%
  \BibitemOpen
  \bibfield  {author} {\bibinfo {author} {\bibfnamefont {Jie-Qiao}\
  \bibnamefont {Liao}}\ and\ \bibinfo {author} {\bibfnamefont {C.~K.}\
  \bibnamefont {Law}},\ }\href@noop {} {\enquote {\bibinfo {title} {Correlated
  two-photon transport in a one-dimensional waveguide side-coupled to a
  nonlinear cavity},}\ } (\bibinfo {year} {2010}),\ \Eprint
  {http://arxiv.org/abs/1009.3335} {arXiv:1009.3335} \BibitemShut {NoStop}%
\bibitem [{\citenamefont {Shi}\ and\ \citenamefont {Sun}(2009)}]{Shi2009}%
  \BibitemOpen
  \bibfield  {author} {\bibinfo {author} {\bibfnamefont {T.}~\bibnamefont
  {Shi}}\ and\ \bibinfo {author} {\bibfnamefont {C.~P.}\ \bibnamefont {Sun}},\
  }\bibfield  {title} {\enquote {\bibinfo {title}
  {{Lehmann-Symanzik-Zimmermann} reduction approach to multiphoton scattering
  in coupled-resonator arrays},}\ }\href {\doibase 10.1103/PhysRevB.79.205111}
  {\bibfield  {journal} {\bibinfo  {journal} {Phys. Rev. B}\ }\textbf {\bibinfo
  {volume} {79}},\ \bibinfo {pages} {205111} (\bibinfo {year}
  {2009})}\BibitemShut {NoStop}%
\bibitem [{\citenamefont {Shi}\ \emph {et~al.}(2010)\citenamefont {Shi},
  \citenamefont {Fan},\ and\ \citenamefont {Sun}}]{Shi2010}%
  \BibitemOpen
  \bibfield  {author} {\bibinfo {author} {\bibfnamefont {T.}~\bibnamefont
  {Shi}}, \bibinfo {author} {\bibfnamefont {S.}~\bibnamefont {Fan}}, \ and\
  \bibinfo {author} {\bibfnamefont {C.~P.}\ \bibnamefont {Sun}},\ }\href@noop
  {} {\enquote {\bibinfo {title} {Two-photon transport in a waveguide coupled
  to a cavity with a two-level system},}\ } (\bibinfo {year} {2010}),\ \Eprint
  {http://arxiv.org/abs/1009.2828} {arXiv:1009.2828} \BibitemShut {NoStop}%
\bibitem [{\citenamefont {Longo}\ \emph {et~al.}(2010)\citenamefont {Longo},
  \citenamefont {Schmitteckert},\ and\ \citenamefont {Busch}}]{Longo2010}%
  \BibitemOpen
  \bibfield  {author} {\bibinfo {author} {\bibfnamefont {P.}~\bibnamefont
  {Longo}}, \bibinfo {author} {\bibfnamefont {P.}~\bibnamefont
  {Schmitteckert}}, \ and\ \bibinfo {author} {\bibfnamefont {K.}~\bibnamefont
  {Busch}},\ }\bibfield  {title} {\enquote {\bibinfo {title} {Few-photon
  transport in low-dimensional systems: Interaction-induced radiation
  trapping},}\ }\href {\doibase 10.1103/PhysRevLett.104.023602} {\bibfield
  {journal} {\bibinfo  {journal} {Phys. Rev. Lett.}\ }\textbf {\bibinfo
  {volume} {104}},\ \bibinfo {pages} {023602} (\bibinfo {year}
  {2010})}\BibitemShut {NoStop}%
\bibitem [{\citenamefont {Gardiner}\ and\ \citenamefont
  {Collett}(1985)}]{Gardiner1985}%
  \BibitemOpen
  \bibfield  {author} {\bibinfo {author} {\bibfnamefont {C.~W.}\ \bibnamefont
  {Gardiner}}\ and\ \bibinfo {author} {\bibfnamefont {M.~J.}\ \bibnamefont
  {Collett}},\ }\bibfield  {title} {\enquote {\bibinfo {title} {Input and
  output in damped quantum systems: Quantum stochastic differential equations
  and the master equation},}\ }\href {\doibase 10.1103/PhysRevA.31.3761}
  {\bibfield  {journal} {\bibinfo  {journal} {Phys. Rev. A}\ }\textbf {\bibinfo
  {volume} {31}},\ \bibinfo {pages} {3761--3774} (\bibinfo {year}
  {1985})}\BibitemShut {NoStop}%
\bibitem [{\citenamefont {Walls}\ and\ \citenamefont
  {Milburn}(2008)}]{Walls2008}%
  \BibitemOpen
  \bibfield  {author} {\bibinfo {author} {\bibfnamefont {D.~F.}\ \bibnamefont
  {Walls}}\ and\ \bibinfo {author} {\bibfnamefont {G.~J.}\ \bibnamefont
  {Milburn}},\ }\href {\doibase 10.1007/978-3-540-28574-8} {\emph {\bibinfo
  {title} {Quantum Optics}}},\ \bibinfo {edition} {2nd}\ ed.\ (\bibinfo
  {publisher} {Springer},\ \bibinfo {year} {2008})\BibitemShut {NoStop}%
\bibitem [{\citenamefont {Scully}\ and\ \citenamefont
  {Zubairy}(1997)}]{Scully2008}%
  \BibitemOpen
  \bibfield  {author} {\bibinfo {author} {\bibfnamefont {M.~O.}\ \bibnamefont
  {Scully}}\ and\ \bibinfo {author} {\bibfnamefont {M.~S.}\ \bibnamefont
  {Zubairy}},\ }\href@noop {} {\emph {\bibinfo {title} {Quantum optics}}}\
  (\bibinfo  {publisher} {Cambridge University Press},\ \bibinfo {year}
  {1997})\BibitemShut {NoStop}%
\bibitem [{\citenamefont {Blow}\ \emph {et~al.}(1990)\citenamefont {Blow},
  \citenamefont {Loudon}, \citenamefont {Phoenix},\ and\ \citenamefont
  {Shepherd}}]{Blow1990}%
  \BibitemOpen
  \bibfield  {author} {\bibinfo {author} {\bibfnamefont {K.~J.}\ \bibnamefont
  {Blow}}, \bibinfo {author} {\bibfnamefont {R.}~\bibnamefont {Loudon}},
  \bibinfo {author} {\bibfnamefont {S.~J.~D.}\ \bibnamefont {Phoenix}}, \ and\
  \bibinfo {author} {\bibfnamefont {T.~J.}\ \bibnamefont {Shepherd}},\
  }\bibfield  {title} {\enquote {\bibinfo {title} {Continuum fields in quantum
  optics},}\ }\href {\doibase 10.1103/PhysRevA.42.4102} {\bibfield  {journal}
  {\bibinfo  {journal} {Phys. Rev. A}\ }\textbf {\bibinfo {volume} {42}},\
  \bibinfo {pages} {4102--4114} (\bibinfo {year} {1990})}\BibitemShut {NoStop}%
\bibitem [{\citenamefont {Loudon}(2000)}]{Loudon2000}%
  \BibitemOpen
  \bibfield  {author} {\bibinfo {author} {\bibfnamefont {R.}~\bibnamefont
  {Loudon}},\ }\href@noop {} {\emph {\bibinfo {title} {The Quantum Theory of
  Light}}},\ \bibinfo {edition} {3rd}\ ed.\ (\bibinfo  {publisher} {Oxford
  University Press},\ \bibinfo {year} {2000})\BibitemShut {NoStop}%
\bibitem [{\citenamefont {Taylor}(2006)}]{Taylor2006}%
  \BibitemOpen
  \bibfield  {author} {\bibinfo {author} {\bibfnamefont {J.~R.}\ \bibnamefont
  {Taylor}},\ }\href@noop {} {\emph {\bibinfo {title} {Scattering theory: the
  quantum theory of nonrelativistic collisions}}}\ (\bibinfo  {publisher}
  {Dover},\ \bibinfo {year} {2006})\BibitemShut {NoStop}%
\bibitem [{\citenamefont {Cushing}(1990)}]{Cushing1990}%
  \BibitemOpen
  \bibfield  {author} {\bibinfo {author} {\bibfnamefont {J.~T.}\ \bibnamefont
  {Cushing}},\ }\href@noop {} {\emph {\bibinfo {title} {Theory construction and
  selection in modern physics: the S matrix}}}\ (\bibinfo  {publisher}
  {Cambridge University Press},\ \bibinfo {year} {1990})\BibitemShut {NoStop}%
\bibitem [{\citenamefont {Goldberger}\ and\ \citenamefont
  {Watson}(1964)}]{Goldberger1964}%
  \BibitemOpen
  \bibfield  {author} {\bibinfo {author} {\bibfnamefont {M.~L.}\ \bibnamefont
  {Goldberger}}\ and\ \bibinfo {author} {\bibfnamefont {K.~M.}\ \bibnamefont
  {Watson}},\ }\href@noop {} {\emph {\bibinfo {title} {Collision Theory}}}\
  (\bibinfo  {publisher} {Wiley},\ \bibinfo {year} {1964})\ pp.\ \bibinfo
  {pages} {209--215}\BibitemShut {NoStop}%
\bibitem [{\citenamefont {Newton}(1982)}]{Newton1982}%
  \BibitemOpen
  \bibfield  {author} {\bibinfo {author} {\bibfnamefont {R.~G.}\ \bibnamefont
  {Newton}},\ }\href@noop {} {\emph {\bibinfo {title} {Scattering Theory of
  Waves and Particles}}},\ \bibinfo {edition} {2nd}\ ed.\ (\bibinfo
  {publisher} {Springer-Verlag},\ \bibinfo {year} {1982})\ pp.\ \bibinfo
  {pages} {156--162}\BibitemShut {NoStop}%
\bibitem [{\citenamefont {Dalton}\ \emph {et~al.}(1999)\citenamefont {Dalton},
  \citenamefont {Barnett},\ and\ \citenamefont {Knight}}]{Dalton1999}%
  \BibitemOpen
  \bibfield  {author} {\bibinfo {author} {\bibfnamefont {B.~J.}\ \bibnamefont
  {Dalton}}, \bibinfo {author} {\bibfnamefont {S.~M.}\ \bibnamefont {Barnett}},
  \ and\ \bibinfo {author} {\bibfnamefont {P.~L.}\ \bibnamefont {Knight}},\
  }\bibfield  {title} {\enquote {\bibinfo {title} {A quantum scattering theory
  approach to quantum-optical measurements},}\ }\href {\doibase
  10.1080/09500349908230404} {\bibfield  {journal} {\bibinfo  {journal}
  {Journal of Modern Optics}\ }\textbf {\bibinfo {volume} {46}},\ \bibinfo
  {pages} {1107--1121} (\bibinfo {year} {1999})}\BibitemShut {NoStop}%
\bibitem [{\citenamefont {Glauber}\ and\ \citenamefont
  {Lewenstein}(1991)}]{Glauber1991}%
  \BibitemOpen
  \bibfield  {author} {\bibinfo {author} {\bibfnamefont {R.~J.}\ \bibnamefont
  {Glauber}}\ and\ \bibinfo {author} {\bibfnamefont {M.}~\bibnamefont
  {Lewenstein}},\ }\bibfield  {title} {\enquote {\bibinfo {title} {Quantum
  optics of dielectric media},}\ }\href {\doibase 10.1103/PhysRevA.43.467}
  {\bibfield  {journal} {\bibinfo  {journal} {Phys. Rev. A}\ }\textbf {\bibinfo
  {volume} {43}},\ \bibinfo {pages} {467--491} (\bibinfo {year}
  {1991})}\BibitemShut {NoStop}%
\bibitem [{\citenamefont {Thompson}\ \emph {et~al.}(1992)\citenamefont
  {Thompson}, \citenamefont {Rempe},\ and\ \citenamefont
  {Kimble}}]{Thompson1992}%
  \BibitemOpen
  \bibfield  {author} {\bibinfo {author} {\bibfnamefont {R.~J.}\ \bibnamefont
  {Thompson}}, \bibinfo {author} {\bibfnamefont {G.}~\bibnamefont {Rempe}}, \
  and\ \bibinfo {author} {\bibfnamefont {H.~J.}\ \bibnamefont {Kimble}},\
  }\bibfield  {title} {\enquote {\bibinfo {title} {Observation of normal-mode
  splitting for an atom in an optical cavity},}\ }\href {\doibase
  10.1103/PhysRevLett.68.1132} {\bibfield  {journal} {\bibinfo  {journal}
  {Phys. Rev. Lett.}\ }\textbf {\bibinfo {volume} {68}},\ \bibinfo {pages}
  {1132--1135} (\bibinfo {year} {1992})}\BibitemShut {NoStop}%
\bibitem [{\citenamefont {Domokos}\ \emph {et~al.}(2002)\citenamefont
  {Domokos}, \citenamefont {Horak},\ and\ \citenamefont
  {Ritsch}}]{Domokos2002}%
  \BibitemOpen
  \bibfield  {author} {\bibinfo {author} {\bibfnamefont {P.}~\bibnamefont
  {Domokos}}, \bibinfo {author} {\bibfnamefont {P.}~\bibnamefont {Horak}}, \
  and\ \bibinfo {author} {\bibfnamefont {H.}~\bibnamefont {Ritsch}},\
  }\bibfield  {title} {\enquote {\bibinfo {title} {Quantum description of
  light-pulse scattering on a single atom in waveguides},}\ }\href {\doibase
  10.1103/PhysRevA.65.033832} {\bibfield  {journal} {\bibinfo  {journal} {Phys.
  Rev. A}\ }\textbf {\bibinfo {volume} {65}},\ \bibinfo {pages} {033832}
  (\bibinfo {year} {2002})}\BibitemShut {NoStop}%
\bibitem [{\citenamefont {Waks}\ and\ \citenamefont
  {Vuckovic}(2006)}]{Waks2006}%
  \BibitemOpen
  \bibfield  {author} {\bibinfo {author} {\bibfnamefont {E.}~\bibnamefont
  {Waks}}\ and\ \bibinfo {author} {\bibfnamefont {J.}~\bibnamefont
  {Vuckovic}},\ }\bibfield  {title} {\enquote {\bibinfo {title} {Dipole induced
  transparency in drop-filter cavity-waveguide systems},}\ }\href {\doibase
  10.1103/PhysRevLett.96.153601} {\bibfield  {journal} {\bibinfo  {journal}
  {Phys. Rev. Lett.}\ }\textbf {\bibinfo {volume} {96}},\ \bibinfo {pages}
  {153601} (\bibinfo {year} {2006})}\BibitemShut {NoStop}%
\bibitem [{\citenamefont {Rephaeli}\ \emph {et~al.}(2010)\citenamefont
  {Rephaeli}, \citenamefont {Shen},\ and\ \citenamefont {Fan}}]{Rephaeli2010}%
  \BibitemOpen
  \bibfield  {author} {\bibinfo {author} {\bibfnamefont {E.}~\bibnamefont
  {Rephaeli}}, \bibinfo {author} {\bibfnamefont {J.~T.}\ \bibnamefont {Shen}},
  \ and\ \bibinfo {author} {\bibfnamefont {S.}~\bibnamefont {Fan}},\ }\bibfield
   {title} {\enquote {\bibinfo {title} {Full inversion of a two-level atom with
  a single-photon pulse in one-dimensional geometries},}\ }\href {\doibase
  10.1103/PhysRevA.82.033804} {\bibfield  {journal} {\bibinfo  {journal} {Phys.
  Rev. A}\ }\textbf {\bibinfo {volume} {82}},\ \bibinfo {pages} {033804}
  (\bibinfo {year} {2010})}\BibitemShut {NoStop}%
\bibitem [{\citenamefont {Gough}\ and\ \citenamefont
  {James}(2009)}]{Gough2009}%
  \BibitemOpen
  \bibfield  {author} {\bibinfo {author} {\bibfnamefont {J.}~\bibnamefont
  {Gough}}\ and\ \bibinfo {author} {\bibfnamefont {M.~R.}\ \bibnamefont
  {James}},\ }\bibfield  {title} {\enquote {\bibinfo {title} {The series
  product and its application to quantum feedforward and feedback networks},}\
  }\href {\doibase 10.1109/TAC.2009.2031205} {\bibfield  {journal} {\bibinfo
  {journal} {Automatic Control, IEEE Transactions on}\ }\textbf {\bibinfo
  {volume} {54}},\ \bibinfo {pages} {2530 --2544} (\bibinfo {year}
  {2009})}\BibitemShut {NoStop}%
\bibitem [{\citenamefont {Kerckhoff}\ \emph {et~al.}(2010)\citenamefont
  {Kerckhoff}, \citenamefont {Nurdin}, \citenamefont {Pavlichin},\ and\
  \citenamefont {Mabuchi}}]{Kerckhoff2010}%
  \BibitemOpen
  \bibfield  {author} {\bibinfo {author} {\bibfnamefont {J.}~\bibnamefont
  {Kerckhoff}}, \bibinfo {author} {\bibfnamefont {H.~I.}\ \bibnamefont
  {Nurdin}}, \bibinfo {author} {\bibfnamefont {D.~S.}\ \bibnamefont
  {Pavlichin}}, \ and\ \bibinfo {author} {\bibfnamefont {H.}~\bibnamefont
  {Mabuchi}},\ }\bibfield  {title} {\enquote {\bibinfo {title} {Designing
  quantum memories with embedded control: Photonic circuits for autonomous
  quantum error correction},}\ }\href {\doibase 10.1103/PhysRevLett.105.040502}
  {\bibfield  {journal} {\bibinfo  {journal} {Phys. Rev. Lett.}\ }\textbf
  {\bibinfo {volume} {105}},\ \bibinfo {pages} {040502} (\bibinfo {year}
  {2010})}\BibitemShut {NoStop}%
\bibitem [{\citenamefont {Haus}(1984)}]{Haus1984}%
  \BibitemOpen
  \bibfield  {author} {\bibinfo {author} {\bibfnamefont {H.~A.}\ \bibnamefont
  {Haus}},\ }\href@noop {} {\emph {\bibinfo {title} {Waves and Fields in
  Optoelectronics}}}\ (\bibinfo  {publisher} {Prentice Hall},\ \bibinfo {year}
  {1984})\BibitemShut {NoStop}%
\end{thebibliography}
\end{document}